\documentclass[
 aip,
 amsmath,amssymb,
 reprint,%
 nofootinbib
]{revtex4-1}

\usepackage{graphicx}
\usepackage{dcolumn}
\usepackage{bm}
\usepackage{enumerate}
\usepackage{siunitx}
\usepackage{amsmath}
\usepackage{cases}
\usepackage{color}

\DeclareSIUnit{\calorie}{cal}

\graphicspath{ {figures/} }

\newcommand{\vecb}[1]{\boldsymbol{\mathrm{#1}}}

\newcommand{\mean}[1]{\left\langle #1 \right\rangle}

\newcommand{\di}{\mathrm{d}} 
\newcommand{\dpart}[2]{\frac{\partial #1}{\partial #2}}
\newcommand{\ddpart}[2]{\frac{\partial^2 #1}{\partial #2^2}}
\newcommand{\deriv}[2]{\frac{\di #1}{\di #2}}

\newcommand{\tp}{{}^\text{T}}

\newcommand{\lB}{\ell_\text{B}}
\newcommand{\kD}{\kappa_\text{D}}
\newcommand{\kB}{k_\text{B}}
\newcommand{\lD}{\lambda_\text{D}}

\newcommand{\rev}[1]{{\color{black} {#1}}}

\newcommand{\ii}{\text{ion-ion}}
\newcommand{\iw}{\text{ion-water}}

\begin{document}


\title{Correlation-induced viscous dissipation in concentrated electrolytes}

\author{Paul Robin}

\email{paul.robin@ista.ac.at}
\affiliation{Laboratoire de Physique de l'\'Ecole Normale Sup\'erieure, ENS, Universit\'e PSL, CNRS, Sorbonne Universit\'e, Universit\'e Paris-Diderot, Sorbonne Paris Cit\'e, Paris, France.}
\affiliation{Institute of Science and Technology Austria, 3400 Klosterneuburg, Austria}

\date{\today}
\begin{abstract}
 	Electrostatic correlations between ions dissolved in water are known to impact their transport properties in numerous ways, from conductivity to ion selectivity. The effects of these correlations on the solvent itself remain, however, much less clear. In particular, the addition of salt has been consistently reported to affect the solution's viscosity -- but most modelling attempts fail to reproduce experimental data even at moderate salt concentration. Here, we use an approach based on stochastic density functional theory, which accurately captures charge fluctuations and correlations. We derive a simple analytical expression for the viscosity correction in concentrated electrolytes, by directly linking it to the liquid's structure factor. Our prediction compares quantitatively to experimental data at all temperatures and all salt concentrations up to the saturation limit. This universal link between microscopic structure and viscosity allows to shed light on the nanoscale dynamics of water and ions in highly concentrated and correlated conditions.
\end{abstract} 
\maketitle

\section{Introduction}

One mole of table salt is dissolved in a liter of pure water: how does this addition modifies the liquid's viscosity? {While this question has been addressed in great details by many experimentalists over the last two centuries \cite{poiseuille1847,sprung1876experimentelle,jones1929viscosity}, their observations are often difficult to rationalize beyond the qualitative level. In particular, the effects of electrostatic interactions between ions dissolved in water are known to be manyfold, with unclear consequences on the liquid's rheological properties}. How these interactions impact ions' transport properties has been the subject of many modelling attempts for over a century, starting with the seminal works of Debye, Hückel, Onsager and others\cite{debye1923theorie,onsager1926theorie,chandra2000ionic}. Most existing theories of electrokinetic transport share, however, many common shortcomings, such as failing at high salt concentrations or for multivalent ions. In addition, the exact nature of the coupling between the motion of dissolved ions and that of the surrounding solvent remains a very much open question, even in the apparently simple case of ions in room temperature water; let alone in more complex environments such as nanoconfined or glass-forming liquids\cite{kavokine2021fluids,robin2023nanofluidics,angell1970glass}, where charge fluctuations can play a key role\cite{robin2023disentangling}.

Here, we focus on the impact of the presence of salt on the liquid's viscosity. It had indeed be noticed first by {Poiseuille \cite{poiseuille1847}} that increasing the salt concentration generally also increases an electrolyte's viscosity, sometimes by up to one order of magnitude near the saturation limit. {Later, Jones and Dole \cite{jones1929viscosity} noted that, in most cases, the relative change of the liquid's viscosity $\eta$ followed an empirical law -- now known as the Jones-Dole equation -- of the form:}
\begin{equation}
	\Delta \eta = \eta - \eta_0 \simeq A \sqrt c + B c,
	\label{eqn:JonesDole}
\end{equation}
where $\eta_0$ is the viscosity of the pure solvent at the same temperature, $c$ is the salt concentration and $A$ and $B$ are empirical, salt- and temperature-dependent parameters. 

\begin{figure}
	\centering
	\includegraphics[width=\linewidth]{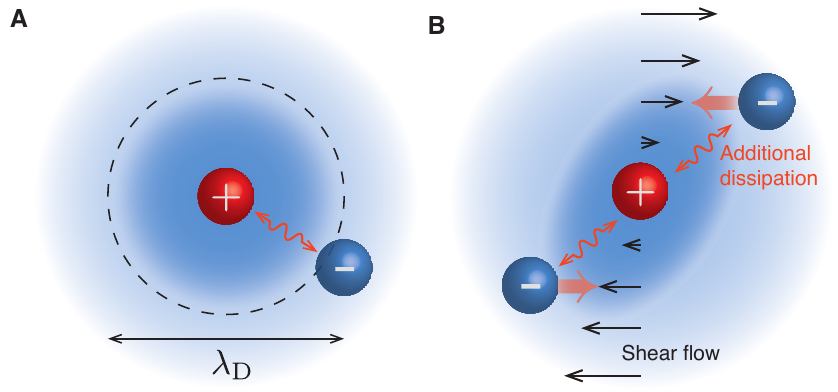}
	\caption{Electrostatic correlations in aqueous electrolytes. \textbf{A} Debye correlation cloud around a cation at thermal equilibrium. \textbf{B} Deformation of the correlation cloud under a shear flow, in the reference frame of a cation. This deformation allows ions to transmit momentum through the fluid via electrostatic interactions, playing the role of an additional viscosity.}
\end{figure}

From the qualitative point of view, the origin of this ``ionic viscosity'' can be readily understood. At thermal equilibrium, ions are typically surrounded by oppositely charged ions, creating a so-called Debye correlation cloud extending over a typical scale known as the Debye length (see Fig. 1A):
\begin{equation}
	\lD = \sqrt{\frac{\epsilon \kB T}{2 e^2 c}},
\end{equation}
where $T$ is temperature, $\epsilon$ the solvent's dielectric constant, $\kB$ Boltzmann's constant and $e$ the elementary charge. In presence of a fluid velocity gradient, however, the correlation cloud is sheared (see Fig. 1B), and electrostatic forces between ions contribute to homogeneize momentum throughout the fluid, effectively increasing its viscosity. Quantifying this effect is a notoriously hard problem, pioneered by Falkenhagen and Debye \cite{falkenhagen1932lxii}. They obtained that, in the limit of infinite dilution, {ion-ion electrostatic interactions are responsible for a viscosity increase of the form:}
\begin{equation}
	\Delta \eta_\ii = \frac{1}{60}\frac{\sqrt{\lB c}}{\sqrt{8 \pi}} \frac{\kB T}{D},
\end{equation}
where we introduced the diffusion coefficient $D$ of ions as well as the Bjerrum length $\lB$, which measures the strength of electrostatic interactions:
\begin{equation}
	\lB = \frac{e^2}{4 \pi \epsilon \kB T}.
	\label{eqn:BjerrumLength}
\end{equation}
This result yields a theoretical prediction for the value of the $A$ coefficient of the Jones-Dole equation \eqref{eqn:JonesDole}. This prediction compares favorably to experiments in the limit of very high dilution (see Fig. 2). 

{At higher concentrations, the $B$ term introduced by Jones and Dole is generally interpreted as describing how individual ions perturb \rev{the solvent -- an effect that is a priori linear in salt concentration. Positive values of $B$ were initially interpreted as stemming from a reinforcement of the hydrogen bond network in water (a phenomenon known as kosmotropy), and conversely for negatives values (chaotropy). Yet, recent works have shown that, while this effect does seem to originate in local electrostatic interactions between ions and water, it does not correspond to large-scale changes in the solvent's structure \cite{gregory2021electrostatic,chialvo2021can}. In addition}, adding this phenomenological term only provides good agreement with experimental data for concentrations up to around 100 mM.
	
Based on these observations, one can write the viscosity increment as the sum of two terms:
\begin{equation}
	\Delta \eta = \Delta \eta_\ii + \Delta \eta_\iw,
\end{equation}
where $\Delta \eta_\ii$ (respectively $\Delta \eta_\iw$) corresponds to the contribution of ion-ion (respectively ion-water) interactions. 
}
\begin{figure}
	\centering
	\includegraphics[width=\linewidth]{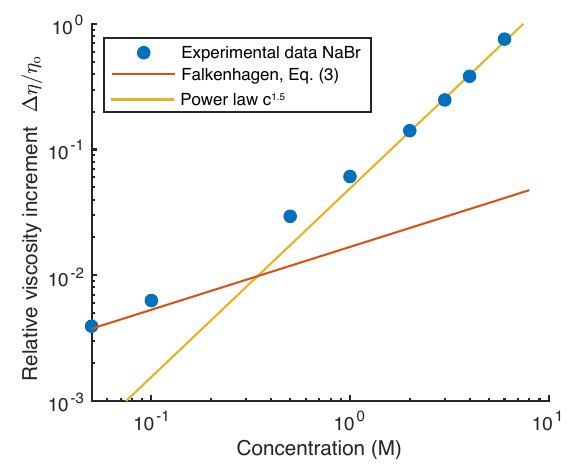}
	\caption{Comparison between experimental data and the Falkenhagen limiting law. Blue points: Experimental data for a NaBr solution at 25$\, \si{\degreeCelsius}$ (reproduced from Ref. \cite{isono1984density}). Red line: Falkenhagen limiting law, $\Delta \eta \propto \sqrt{c}$. Yellow line: power law fit $\Delta \eta \propto c^{1.5}$ in the limit of high concentrations.}
\end{figure}

{ Various attempts at extending the Jones-Dole law were reported in the literature  \cite{jones1940viscosity,onsager2002irreversible,esteves2001debye}; they generally amount to adding phenomenological terms scaling as $c^2$, $c \log c$, etc., emerging from e.g. volume exclusion effects, ion-ion interactions or electrostatic barriers for microscopic rearrangements -- without strong theoretical evidence for any of the suggested scalings. 
In addition, the suggested models contain various fitting parameters that do not allow for easy physical interpretation, or only compare reasonably to certain salts or experimental conditions. As an example of such limitations, in the limit of very high concentrations (above 1 M), the viscosity increment seems to scale like $c^{3/2}$, see Fig. 2 in the case of sodium bromide (NaBr). This scaling differs from the ones often used in the literature to extend the Jones-Dole law.
}

\begin{figure}
	\centering
	\includegraphics[width=\linewidth]{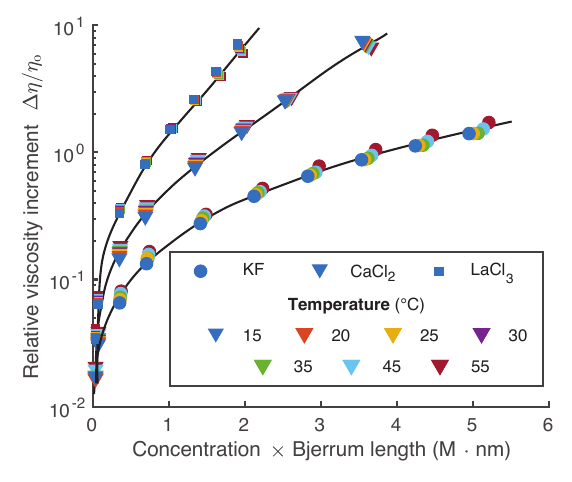}
	\caption{Rescaled viscosity increment as function of concentration and temperature. $x$ axis: $c \times \lB$ (note that $\lB$ depends on temperature); $y$ axis: $\Delta \eta/\eta_0(T)$. Symbols represent experimental data for KF, CaCl$_2$ and LaCl$_3$, reproduced from Refs. \cite{goldsack1978viscosity,isono1984density}. Colors represent temperatures. Black lines are guides for the eye.}
\end{figure}

{It is not a priori clear whether these high-concentration deviations arise from ion-ion or ion-water interactions. However, rescaling the viscosity increment by the viscosity of pure water $\eta_0(T)$, and 
plotting it as function of the quantity $c \lB$, experimental data for a given salt at all temperatures seem to collapse on a single mastercurve at high concentrations. Since $c \lB \propto \lD^{-2}$ is a measure of electrostatic correlations between ions, this observation suggests that high concentration deviations arise mostly from ion-ion interactions.
}

Based on these observations, we develop in this paper a field-theoretical approach for the description of this ionic viscosity, and show that it can be directly determined from the charge structure factor of the electrolyte. The latter can be determined thanks to a stochastic density functional theory based on the Dean-Kawasaki equation\cite{kawasaki1994stochastic,dean1996langevin}, which has recently proved successful at determining various properties of electrolytes \cite{demery2016conductivity,avni2022conductivity,avni2022conductance,bernard2023analytical}. In particular, we make use of a technique introduced by Avni and coworkers\cite{avni2022conductivity} to cut off electrostatic interactions at short distance, enabling to better describe the structure of concentrated electrolytes. 

Overall, we show as our main result that the viscosity increment can determined analytically as a Fourier space integral:
{
\begin{equation}
	\Delta \eta_\ii  =  \frac{c }{15(2 \pi)^2}\frac{\kB T}{D} \int \di  q  V( q) \deriv{}{q} \left[ q^2 C_\rho^0(q) \dpart{C_\rho^0}{q} \right],
\end{equation}
}
where $V(q)$ is the Fourier-transformed ion-ion interaction potential (e.g. electrostatic or van der Waals interactions), and $C_\rho^0$ is the charge structure factor of the electrolyte at thermal equilibrium in absence of any flow -- directly linking the electrolyte's microscopic structure to a macroscopic quantity such as viscosity. Based on this result, a simple ansatz for the viscosity of concentrated electrolytes is found to be:
{
\begin{equation}
	\Delta \eta =  B(T) c + \frac{1}{60 \sqrt{8 \pi }} \frac{\kB T}{D}  \left[\sqrt{\lB c } + 6 \pi a^2 \left(  \lB c\right)^{3/2}\right],
	\label{eqn:Main}
\end{equation}
where the only new parameter introduced by our model is the ionic size $a$, allowing for straightforward physical interpretation. }We show that this theoretical result compares favorably to experimental data in nearly all conditions of temperature, concentration, salt composition and valence, and can correctly predict viscosity increments of more than $300 \%$, in the case of multivalent salts close to the saturation limit.

This paper is organized as follows. In Section II, we present our field-theoretical framework and derive equation \eqref{eqn:Main}. Readers not interested in the details of the computation may skip Sections \ref{ss:DK} to \ref{ss:Mult} and go directly to Section III, where we compare our predictions to a large body of experimental data. We also provide a simple, quasi-quantitative interpretation of our result. Finally, Section IV establishes our conclusion.

\section{From electrostatic correlations to the ionic viscosity}

\subsection{Hydrodynamics with ions}

We consider an aqueous solution containing some monovalent binary salt $X^+, Y^-$ with concentration $c$, subjected to a shear flow. {We only consider ion-ion interactions, and therefore treat water as a continuous fluid with a given permittivity $\epsilon(T)$.} {We assume that both types of ions are monovalent with same diffusion coefficient $D$ and have the same physical size; the relaxation of these assumptions will be discussed later. We denote by $\vecb u(\vecb r)$ the local fluid velocity, $p(\vecb r)$ the pressure field, $n_+(\vecb r)$ the local density in cations and $n_-(\vecb r)$ the local density in anions. We define $\rho (\vecb r) = n_+(\vecb r) - n_-(\vecb r)$ the local charge density. It should be noted that $\mean{\rho} = 0$ due to electroneutrality and that $\mean{n_+} = \mean{n_-} = c$}. In the limit of low Reynolds numbers, the velocity field solves the Stokes equation:
\begin{align}
	\eta_0 \vecb \nabla^2 \vecb u - \vecb \nabla  p + \vecb f(\vecb r) = 0,
\end{align}
where $\vecb f$ represents all {body} forces acting on the fluid, other than pressure. If the fluid is subject to no external force, then $\vecb f(\vecb r)$ corresponds only to interactions between dissolved ions at position $\vecb r$ and other ions elsewhere, which derives from an interaction potential $V(r)$:
\begin{equation}
	\vecb f(\vecb r) =- \kB T \int \di \vecb r' \rho(\vecb r) \vecb \nabla V(\vecb r - \vecb r') \rho(\vecb r').
\end{equation}
{In the simplest case, where we assume that ions are point particles with no short-range repulsion, the interaction potential is simply the  Coulomb potential: $V (\vecb r) = e^2/4 \pi \kB T \epsilon r = \lB/r$. Other situations will also be addressed later}. Since the fluid is electroneutral on average, $\mean{\rho} = 0$ and the net force $\mean{\vecb f}$ acting on the fluid is solely due to local and random charge fluctuations. Introducing the Fourier transform as:
\begin{equation}
	\vecb f(\vecb k) = \int \vecb f (\vecb r) e^{i \vecb k \cdot \vecb r} \, \di \vecb r,
\end{equation}
we can express the electrostatic force as:
{
\begin{equation}
	\mean{\vecb f(\vecb k)} = \frac{\kB T}{(2 \pi)^3} \int \di \vecb q \mean{\rho(\vecb k - \vecb q) \rho(\vecb q)} (i \vecb q) V(q).
	\label{eqn:f-gamma}
\end{equation}
}
This force vanishes at equilibrium due to the problem's symmetries, but may take a non-zero value in presence of an external shear flow. The last equation allows to directly link $\mean {\vecb f}$ to the electrolyte's charge structure factor defined as:
\begin{equation}
	\Gamma_\rho (\vecb k, \vecb k') = \mean{\rho(\vecb k) \rho(\vecb k')}.
\end{equation}
The goal of next sections is therefore to compute this structure factor using a field-theoretical approach, and to use equation \eqref{eqn:f-gamma} to show that 
\begin{equation}
	\mean{\vecb f(\vecb k)} \simeq - k^2 \Delta \eta \vecb u(\vecb k).
	\label{eqn:DeltaEta}
\end{equation}
To do so, we first compute the structure factor $\Gamma_\rho$ at equilibrium (i.e. in the absence of any flow), and use it to compute the effect of advection by the solvent when an external flow is present. We then deduce a first-order correction of the structure factor, and use equation \eqref{eqn:f-gamma} to obtain the viscosity correction. We now present the details of this computation.

\subsection{Charge fluctuations and the Dean-Kawasaki equation}
\label{ss:DK}

The local charge density $\rho$ can be determined by computing the fluctuations of the local cation and anion densities, {$n_+$ and $n_-$,} around their mean value $c$. These fluctuations can be described by the Dean-Kawasaki equation\cite{kawasaki1994stochastic,dean1996langevin}, which has been used recently to compute the conductivity of concentrated electrolytes accurately \cite{avni2022conductance}. {It reads:}
\begin{widetext}
\begin{equation}
	\partial_t n_\pm = - \vecb \nabla \cdot (n_\pm \vecb u ) + D \nabla^2 n_\pm \pm D \vecb \nabla \cdot  \int \di \vecb r' n_\pm(\vecb r)  \vecb \nabla V(\vecb r - \vecb r') \rho(\vecb r') + \sqrt{2 D n_\pm (\vecb r)} \vecb \nabla \cdot \vecb \zeta_\pm,
	\label{eqn:DK}
\end{equation}
where {$\zeta_+$ and $\zeta_-$ are} uncorrelated white noise fluxes with zero mean and unit variance. The first term on the right-hand side corresponds to advection by the solvent, the second to ion diffusion, the third to ion-ion interactions and the last one to random Brownian fluctuations.

In what follows, we assume that fluctuations of  {$n_+$ and $n_-$ are} small compared to the average value $c$, so that we may work at first order in $\delta n_\pm = n_\pm - c$ and $\zeta_\pm$. {Since $\rho = n_+ - n_-$} and that $\vecb \nabla \cdot \vecb u = 0$ due to the fluid being incompressible, we obtain:
\begin{equation}
	\partial_t \rho = - \vecb u \cdot \vecb \nabla \rho + D \nabla^2 \rho + 2 c D \vecb \nabla \cdot  \int \di \vecb r'  \vecb \nabla V(\vecb r - \vecb r') \rho(\vecb r') + \sqrt{4 D c} \vecb \nabla \cdot \vecb \zeta,
\end{equation}
or, in Fourier space:
\begin{equation}
	\partial_t \rho(\vecb k) = \frac{1}{(2 \pi)^3}\int \di \vecb q\,  \vecb u(\vecb k- \vecb q) \cdot i \vecb q \rho(\vecb q) - D k^2 \rho(\vecb k) - D \kD^2 \rho(\vecb k) - i \sqrt{4 D c} k \zeta.
	\label{eqn:DKFourier}
\end{equation}
Note that we made use of the fact that the sum of two Gaussian vectors is itself a Gaussian vector, with additive variance{, and that $V(q) = 4 \pi \lB/q^2$ in the case of point-like ions}. We also introduced the inverse Debye length $\kD = \lD^{-1}$. 

The last equation can be seen as an evolution equation of the form:
\begin{equation}
	\partial_t \vecb \rho = \left(- \vecb a + i \mathcal L \right) \cdot \vecb \rho + \vecb b ,
	\label{eqn:Zwanzig}
\end{equation}
where $\vecb \rho$ is the vector $\left\{\rho(\vecb k)\right\}_{\vecb k}$, and where the operators $\vecb a$, $\vecb b$ and $\mathcal L$ are given by:
\begin{align}
	\vecb a &= \text{diag}\left\{ D k^2 + D \kD^2 \right\}_{\vecb k},\\
	\vecb b &= \left\{ - i \sqrt{4Dc} k \zeta (\vecb k)\right\}_{\vecb k},\\
	\mathcal L \cdot f(\vecb k, \vecb k') &=  \frac{1}{(2 \pi)^3}\int \di \vecb q\,  \vecb u(\vecb k- \vecb q) \cdot \vecb q f(\vecb q, \vecb k').
\end{align}
We may now integrate equation \eqref{eqn:DKFourier} over time, assuming that the initial condition vanishes:
\begin{equation}
	\rho(\vecb k, t) = \left[\int_0^t e^{(- \vecb a + i \mathcal L) (t-s)} \cdot \vecb b\, \di s \right]_{\vecb k}.
\end{equation}
The charge structure factor can now be obtained as:
\begin{equation}
	\Gamma_\rho(\vecb k, \vecb k';t) = \mean{\vecb \rho(t) \cdot \vecb \rho^\dagger(t)}_{\vecb k \vecb k'} = \int_0^t \di s \int_0^t \di s' e^{(- \vecb a + i \mathcal L) (t-s)} \cdot \mean {\vecb b \cdot \vecb b^\dagger } \cdot e^{(- \vecb a - i \mathcal L \tp ) (t-s')} .
\end{equation}
Since we are interested in static correlations, we now take the limit $t\to \infty$ and use the fact that:
\begin{equation}
	\mean{\zeta(\vecb k, t) \zeta(\vecb k', t')} = (2 \pi)^3 \delta(\vecb k + \vecb k') \delta(t - t').
\end{equation}
We obtain (see Ref. \cite{zwanzig2001nonequilibrium}):
\begin{equation}
	(- \vecb a + i \mathcal L) \cdot \Gamma_\rho + \Gamma_\rho \cdot (- \vecb a - i \mathcal L \tp ) = - 4 D c k^2 (2 \pi)^3 \delta(\vecb k + \vecb k'),
	\label{eqn:ZwanzigSol}
\end{equation}
or, using the definition of $\mathcal L$ and $\mathbf a$:
\begin{equation}
	D(k^2 + k'^2 + 2 \kD^2)\Gamma_\rho(\vecb k, \vecb k')- \frac{i}{(2 \pi)^3} \int \di \vecb q\,  \vecb u(\vecb k- \vecb q) \cdot \vecb q \Gamma_\rho(\vecb q, \vecb k') + \frac{i}{(2 \pi)^3} \int \di \vecb q\,  \vecb u(\vecb k'- \vecb q) \cdot \vecb q \Gamma_\rho(\vecb k, \vecb q) = 4 D c k^2(2 \pi)^3 \delta(\vecb k + \vecb k').
	\label{eqn:Convol}
\end{equation}
Equation \eqref{eqn:Convol} cannot be solved directly. However, since we only wish to compute $\mean{\vecb f}$ up to linear order in the velocity field $\vecb u$, one may use a perturbative approach, which we now describe. At zeroth order in $\vecb u$, one may set $\mathcal L = 0$ and obtain:
\begin{align}
	\Gamma_\rho^0(\vecb k , \vecb k') = (2 \pi)^3 \delta(\vecb k + \vecb k' )\frac{2 c k^2}{k^2 + \kD^2} = (2 \pi)^3 \delta(\vecb k + \vecb k' ) g(k).
	\label{eqn:Gamma0}
\end{align}
To obtain the correction at first order in $\vecb u$, one may simply replace $\Gamma_\rho$ by $\Gamma_\rho^0$ in all convolutions in equation \eqref{eqn:Convol}. We obtain:
\begin{equation}
	\Gamma_\rho(\vecb k, \vecb k') = \Gamma_\rho^0(\vecb k , \vecb k') - \frac{\vecb u (\vecb k + \vecb k') \cdot(i \vecb k' g(k') - i \vecb k g(k))}{D (k^2 + k'^2 + 2 \kD^2)}.
\end{equation}
Inserting this result in equation \eqref{eqn:f-gamma} then yields:
\begin{equation}
	\mean{\vecb f(\vecb k)}  = \frac{\kB T}{(2 \pi)^2D} 2 \lB \int \di \vecb q \frac{g(q) - g(\vecb k- \vecb q)}{q^2 + (\vecb k- \vecb q)^2 + 2 \kD^2} \frac{\vecb q \vecb q}{q^2}  \cdot \vecb u(\vecb k).
\end{equation}
Note that we have used the incompressibility condition $\vecb u (\vecb k) \cdot \vecb k = 0$. To compute this integral, we use spherical coordinates $(q, \theta, \phi)$ {to describe the Fourier space}, with $\vecb k $ being the reference for angles. We have:
\begin{equation}
	\frac{\vecb q \vecb q}{q^2} =
	\begin{pmatrix}
		\cos^2 \theta & \sin \theta \cos \theta \cos \phi & \sin \theta \cos \theta \sin \phi \\
		\sin \theta \cos \theta \cos \phi & \sin^2 \theta \cos^2 \phi & \sin^2 \theta \sin \phi \cos \phi \\
		\sin \theta \cos \theta \sin \phi & \sin^2 \theta \sin \phi \cos \phi & \sin^2 \theta \sin^2 \phi
	\end{pmatrix}.
\end{equation}
The rest of the integrand does not depend on $\phi$, so we can write:
\begin{equation}
	\int_0^{2 \pi} \di \phi \frac{\vecb q \vecb q}{q^2} =
	\pi	\begin{pmatrix}
		2\cos^2 \theta & 0& 0 \\
		0 & \sin^2 \theta  & 0 \\
		0 & 0 & \sin^2 \theta
	\end{pmatrix}.
\end{equation} 
Note that the first diagonal coefficient is irrelevant since we only wish to evaluate the tensor on a vectorial subspace orthogonal to $\vecb k$ due to incompressibility. Since we interpret $\vecb f$ as a viscous force, we are interested in its long wavelength limit $k \to 0$ (see equation \eqref{eqn:DeltaEta}). Expanding to second order in $k$ and integrating over $\theta$, we obtain:
\begin{equation}
	\mean{\vecb f (\vecb k)} \simeq - \frac{\kB T}{D} \lB \frac{2 }{15 \pi} c \int_0^\infty \di q \kD^2 q^2 \frac{5 \kD^2 - q^2}{(\kD^2 + q^2)^4}k^2 \vecb u (\vecb k) = - \frac{\kB T}{D} \lB c \frac{1}{60 \kD} k^2 \vecb u (\vecb k).
\end{equation}

Comparing with equation \eqref{eqn:DeltaEta}, we find that indeed electrostatic interactions between ions effectively increase the fluid's viscosity by:
{
\begin{equation}
	\Delta \eta_\ii =  \frac{1}{60 \sqrt{8 \pi }} \frac{\kB T}{D}  \sqrt{\lB c },
	\label{eqn:Falkenhagen}
\end{equation}
}
which corresponds to the Falkenhagen limiting law \cite{falkenhagen1932lxii}.

\subsection{General case: universal link between the charge structure factor and the viscosity increment}
The above Fourier-space approach has the advantage of being computationally lighter than Falkenhagen's historical real-space derivation (which made extensive use of spherical harmonics), and of offering a straightforward way of extending the result to any desired accuracy, provided that the charge structure factor in absence of flow $\Gamma_\rho^0$ is known, and might deviate from equation \eqref{eqn:Gamma0} due to non-Coulombic interactions between ions, e.g. short-distance repulsion. In the case of a generic interaction potential $V(\vecb k)$, due to translation invariance in the absence of external flow, $\Gamma_\rho^0$ can always be written as:
\begin{equation}
	\Gamma_\rho^0(\vecb k, \vecb k') = 2 c (2 \pi)^3 \delta(\vecb k + \vecb k')C_\rho^0(\vecb k),
\end{equation}
where we introduced the rescaled structure factor $C_\rho^0(k)$. One can then express the viscosity increment as function of this quantity alone:
\begin{equation}
	\Delta \eta  =  \lim\limits_{k\to 0}\frac{1}{(2 \pi)^3 k^2}\frac{\kB T}{D}  \int \di \vecb q V(\vecb q) \frac{C_\rho^0(\vecb q) - C_\rho^0(\vecb k- \vecb q)}{(\vecb k - \vecb q)^2C_\rho^0(\vecb q) + q^2C_\rho^0(\vecb k- \vecb q)} C_\rho^0(\vecb q) C_\rho^0(\vecb k- \vecb q) \vecb q \vecb q.
\end{equation}
{If we assume ions to be perfectly spherical, then $C_\rho^0(\vecb k)$ only depends on $k$. This assumption is valid for atomic ions (Na$^+$, Cl$^-$, etc.), and is an approximation in the case of molecular ions (SO$_4^{2-}$, NO$_3^-$, etc.). We can therefore expand the integrand for $k \to 0$ and perform the integral of $\phi$ and $\theta$, yielding:
}
{
\begin{equation}
	\boxed{	\Delta \eta_\ii  =  \frac{c }{15(2 \pi)^2}\frac{\kB T}{D} \int \di  q  V( q) \deriv{}{q} \left[ q^2 C_\rho^0(q) \dpart{C_\rho^0}{q} \right].}
	\label{eqn:DeltaEtaGeneral}
\end{equation}
}
\end{widetext}

In the above case where ions interact through eletrostatics alone, {with no short-range repulsion,} then the rescaled structure factor is given by
\begin{equation}
	C_\rho^0 (k) = \frac{k^2}{k^2 + \kD^2},
	\label{eqn:C_DH}
\end{equation}
and equation \eqref{eqn:DeltaEtaGeneral} is equivalent to equations \eqref{eqn:DeltaEta} and \eqref{eqn:Falkenhagen}. However, equation \eqref{eqn:C_DH} only provides a very rough estimate of the electrolyte's structure factor, which can be determined from numerical simulations or experiments. For example, Fig. 4A shows a comparison between equation \eqref{eqn:C_DH} (red line) and results from molecular dynamics simulations of a concentrated NaCl solution \cite{kim2023electrical} (blue circles). The above ansatz fails at capturing the layered structure of ionic correlations at high concentrations, as those emerge from short-distance repulsions between ions.

Consequently, a straightforward way of improving on Falkenhagen's result is to use a more precise ansatz for $C_\rho^0(k)$, and insert it into equation \eqref{eqn:DeltaEtaGeneral}. Although this ansatz does not need to be physically motivated as long as it faithfully reproduces experimental and numerical data, we report results to this end in next section, making use of a simple model first introduced by Ref. \cite{avni2022conductivity}.

\subsection{Finite ion size and truncated Coulomb potential}

\begin{figure*}
	\centering
	\includegraphics[width=0.7\linewidth]{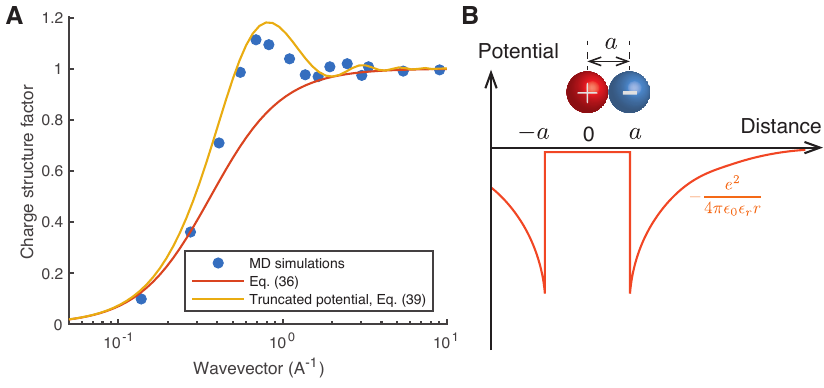}
	\caption{Effective cut-off electrostatic potential. {\textbf{A} Charge structure factor of a concentrated electrolyte. Blue points: molecular dynamics simulation of a solution of NaCl with concentration $c = 1.23 \,$M (adapted from Ref. \cite{kim2023electrical}). Red line: Equation \eqref{eqn:Gamma0} (no cut-off). Yellow line: Equation \eqref{eqn:Gamma0_Avni} (cut-off $a = 3\, \si{\angstrom}$). \textbf{B} Definition of the short-distance cut-off $a$. $a$ corresponds roughly to the distance of minimal approach between two ions: instead of introducing a short-distance repulsion, the interaction potential is simply truncated below $a$.}}
\end{figure*}

Introducing a short-distance repulsion between ions in the Dean-Kawasaki equation unfortunately makes the computation intractable. However, Avni and coworkers suggested an alternative approach to account for the finite size of ions \cite{avni2022conductivity}: truncating the Coulomb potential at some finite cut-off distance $a$, and setting the ion-ion interaction to zero below $a$ (see Fig. 4B):
\begin{equation}
	V(r) = \frac{\lB}{r} \to \frac{\lB}{r} H(r-a),
\end{equation}
with $H$ the Heaviside step function. In Fourier space, this corresponds to:
\begin{equation}
	V(k) = 4 \pi \frac{\lB}{k^2} \to 4 \pi \frac{\lB}{k^2} \cos k a.
	\label{eqn:TruncFour}
\end{equation}
While this constitutes an apparently uncontrolled approximation, this trick generally yields accurate results, at very little mathematical cost. 

Overall, the above derivation of $\Gamma_\rho^0$ still holds, replacing $\kD^2$ by $\kD^2 \cos k a$ in equation \eqref{eqn:DKFourier}. The equilibrium structure factor $C_\rho^0$ is then given by:
\begin{equation}
	C_\rho^0(k) = \frac{k^2}{\kD^2 \cos k a+ k^2}.
	\label{eqn:Gamma0_Avni}
\end{equation}
This result is shown on Fig. 4A (yellow line), and captures the essential features of the numerical data. Therefore, truncating the electrostatic potential appears to a be viable strategy to accurately describe the structure of concentrated electrolytes.

Injecting now equation \eqref{eqn:Gamma0_Avni} in \eqref{eqn:DeltaEtaGeneral}, we obtain at leading order in $a$:
\begin{equation}
	\boxed{\Delta \eta_\ii =  \frac{1}{60 \sqrt{8 \pi }} \frac{\kB T}{D}  \left[\sqrt{\lB c } + 6 \pi a^2 \left(  \lB c\right)^{3/2}\right].}
	\label{eqn:Correction}
\end{equation}
Equations \eqref{eqn:DeltaEtaGeneral} and \eqref{eqn:Correction} constitute the main result of this work.

\subsection{Multivalent ions and asymmetric salts}
\label{ss:Mult}

In the previous sections, we only considered the case of a monovalent binary electrolyte $X^+, Y^-$. However, the discussion can be extented in a straightforward manner to multivalent salts of the type $z:1$ or $z:z$ with $z > 1$. Noticing that $\rho = z_+ n_+ - z_- n_-$, all the above derivations can be redone, yielding:
 \begin{equation}
 	\Delta \eta =  \frac{1}{60 \sqrt{8 \pi }} \frac{\kB T}{D}  \left[\sqrt{z \bar z\lB c } + 6 \pi a^2 \left( z \bar z \lB c\right)^{3/2}\right],
 	\label{eqn:Multivalent}
 \end{equation}
with $\bar z = z$ or $(z+1)/2$ depending for $z:z$ and $z:1$ salts, respectively. 

{In a similar manner, the case where cations and anions have different diffusion coefficients, say $D_+$ and $D_-$, can also be treated analytically, see Refs. \cite{falkenhagen1932lxii} and \cite{avni2022conductance}. It can be shown that all the above computations still hold, deplacing the quantity $D$ by an effective coefficient:
	\begin{widetext}
\begin{equation}
	D_\text{eff} = \frac{(z_+ + z_-) D_+ D_-}{D_+  z_- + D_- z_+ - 4 z_+ z_- \left(\frac{D_+ - D_-}{\sqrt{D_+ + D_-} \sqrt{z_+ + z_-} + \sqrt{D_+ z_+ + D_- z_-}} \right)^2}
	\label{eqn:DiffCoef}
\end{equation}
\end{widetext}

In the above derivation, we also used the same cut-off distance $a$ for cation-cation, cation-anion and anion-anion interactions. One could, in principle, define specific values of a for cation-cation and anion-anion interactions, and then use mixing rules for cross-interactions. Doing so has been reported to only marginally affect the results \cite{avni2022conductance}; we therefore use a single value of $a$ for all types of interactions.

\subsection{Discussion of the cut-off potential}

\label{SecIIF}
Lastly, let us comment the choice of the cut-off potential. While the combination of the cut-off potential and of the stochastic density functional theory has been shown to be quite powerful to account for finite size effects in the transport dynamics of concentrated electrolytes \cite{avni2022conductivity}, recent developments \cite{bernard2023analytical} have suggested that details of the cut-off potential may need to be chosen carefully. In particular, Ref. \cite{bernard2023analytical} suggests to set the potential to some finite value $v_0$ (which may be positive or negative) at distances smaller than $a$:
\begin{equation}
	V(r) = \frac{\lB}{r} H(r-a) + v_0 H(a-r).
\end{equation}
In this case, it can be shown that the ion-ion correlator becomes:
\begin{equation}
	C_\rho^0(k) = \frac{k^2}{k^2 + \kD^2 \cos k a + v_0 \kD^2 (\sin k a - ka \cos ka)/\lB k}.
	\label{eqn:Gamma0_Illien}
\end{equation}
Inserting again this result into equation \eqref{eqn:Main}, we obtain:
\begin{equation}
	\Delta \eta =  \frac{\kB T \sqrt{\lB c } }{60 D \sqrt{8 \pi }} \left[1 + 6 \pi a^2 \left[1 - \frac 2 3 \frac{a v_0}{\lB}\right] \lB c\right],
	\label{eqn:CorrectionIllien}
\end{equation}
which is identical to the previous result, with $a$ being effectively replaced by $a \sqrt{1 - 2av_0/3 \lB} $. 

Since we can a priori expect $v_0$ to be at most of the order of $\lB/a$, this modification essentially amounts to modifying $a$ by a factor of order unity. Our results should therefore not depend too much on the exact details of the cut-off potential. In what follows, we come back to the simple case where $v_0 = 0$, and instead treat $a$ as an adjustable parameter.
}

\section{Interpretation and comparison with experimental data}

\subsection{Physics of the ionic viscosity and the truncated potential}

At the semi-quantitative level, one can interpret the Falkenhagen limiting law (equation \eqref{eqn:Falkenhagen}) and the existence of a viscosity increment at low concentrations as follows. As previously stated, ions in an electrolyte at equilibrium are typically surrounded by a Debye atmosphere bearing an opposite charge, and distributed over a typical lengthscale $\lD$.

Let us now consider the case where an external flow is applied on the electrolyte, with a given velocity gradient $\dpart u z$, see Fig. 1B. We notice that, since ions in the correlation cloud are typically separated by $\lD$, they will feel different solvent velocities, typically by an amount $\lD \dpart u z(z)$, where $\dpart u z$ is the external velocity gradient. Therefore, an anion in the Debye cloud surrounding a cation will be on average pulled away by a force $\lD \dpart u z/\mu$, where $\mu$ is the ion's mobility. The energy landscape of the Debye atmosphere is locally modified by $\Delta E \sim \lD^2 \dpart u z / \mu$ upstream of the flow and by $-\Delta E$ downstream: the probabilities of finding an anion there are modified by $e^{\pm \Delta E/ \kB T}$, tilting the cloud along the velocity profile (see Fig. 1B). Since each ion in the correlation cloud is exerting an average force $e^2/4 \pi \epsilon \lD^2$ on the central cation, the latter overall feels a net force of the order of:
\begin{equation}
	\frac {\lD^2\lB} {\lD^2 \mu} \left( \dpart u z (+\lD) - \dpart u z (- \lD)\right) \sim \frac {\lD\lB} {\mu}\ddpart{v}{z}.
\end{equation}
Since the electrolyte has a concentration $c$, the overall force exerted on the liquid is
\begin{equation}
	f \sim \frac{c \lB \lD}{\mu} \ddpart u z,
\end{equation}
which shows that the presence of ions is equivalent to an additional viscosity of the order of 
\begin{equation}
	\Delta \eta \sim \frac{c \lB \lD}{\mu} \sim \frac{1}{\mu} \sqrt{\lB c}.
\end{equation}
Importantly, this simple argument explains why this correction scales like the inverse of the mobility $\mu$, and identifies the quantity $\lB c$ as the main relevant parameter.

Furthermore, the use of the truncated potential \eqref{eqn:TruncFour} can be justified from the theoretical point of view by comparing this ansatz to the so-called Poisson-Fermi equation introduced to account for crowding effects in concentrated electrolytes \cite{bazant2011double}:
\newcommand{\lC}{\ell_\text{c}}
\begin{equation}
	(1 - \lC^2 \nabla^2)\nabla^2 V = - 4 \pi \lB \rho,
	\label{eqn:PoissonFermi}
\end{equation}
where $\lC$ is a measure of the ionic size. Solving equation \eqref{eqn:PoissonFermi} around a point-like charge $\rho = \delta(\vecb r)$ and Fourier transforming yields:
\begin{equation}
	V(k) = \frac{\lB}{k^2} \frac{4 \pi }{1 + \lC^2 k^2} \simeq 4 \pi \frac{\lB}{k^2} \left[ 1 - \lC^2 k^2\right] \simeq 4 \pi \frac{\lB}{k^2} \cos 2 k \lC,
\end{equation}
which corresponds to equation \eqref{eqn:TruncFour} with $a = 2 \lC$, strengthening our otherwise uncontrolled approximation.

Lastly, we can interpret the fact that truncating the potential actually results in a larger viscosity correction. The charge structure factor contains terms corresponding to cation-cation, anion-anion and cation-anion correlations. If no interaction cut-off nor short-distance repulsion are introduced, then nothing prevents oppositely charge ions to significantly overlap each other, being separated by $\lD$ which can become smaller than $a$ at high concentrations. If ions overlap, they essentially form a neutral pair that does not interact with the environment, and becomes ineffective at transmitting momentum over large distance. Instead, if ions cannot be closer than some finite distance $a$, then interactions are not entirely screened off at high concentration and the resulting ionic viscosity continues to increase sharply.
\begin{figure*}
	\centering
	\includegraphics[width=\linewidth]{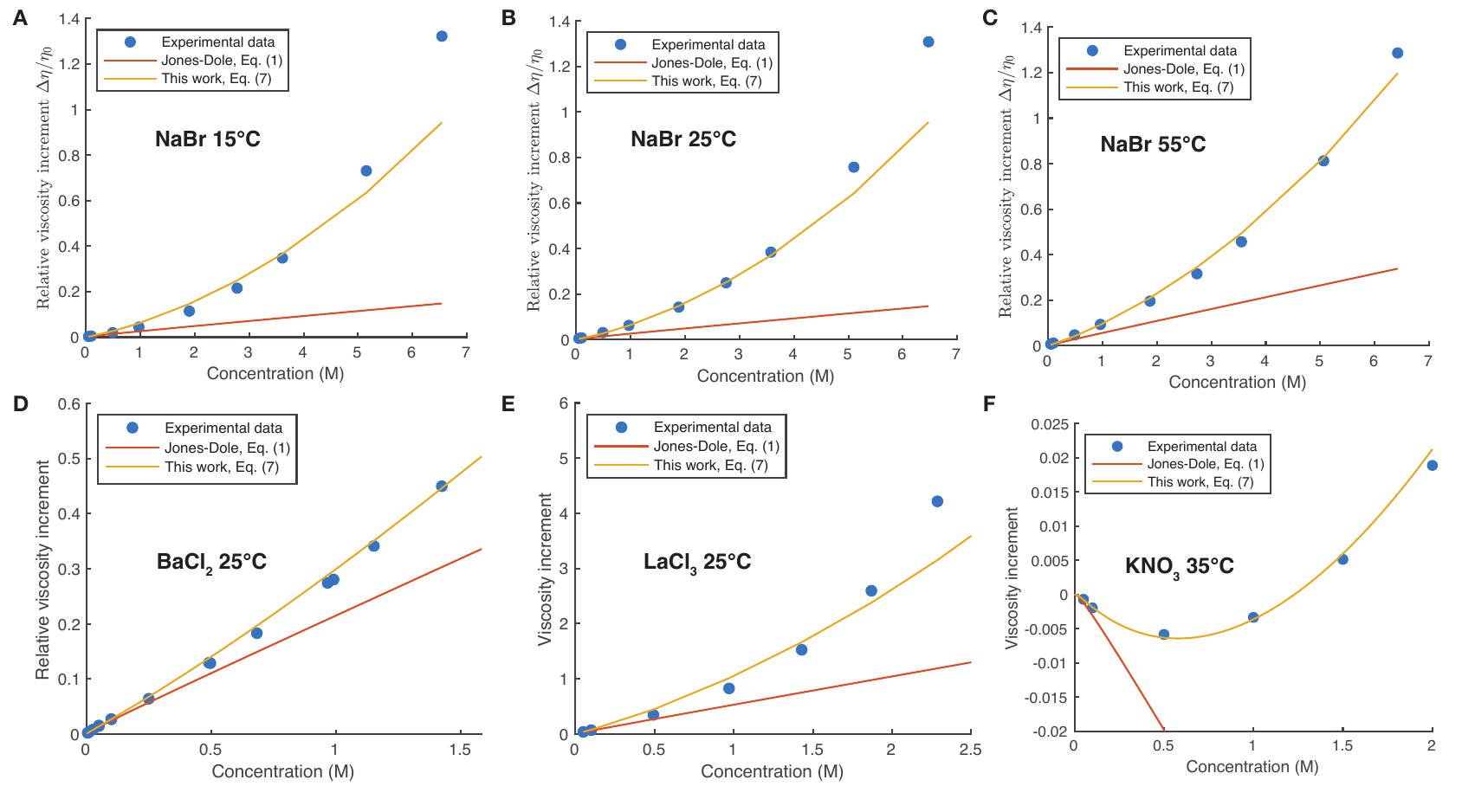}
	\caption{Comparison between theory (equation \eqref{eqn:Main}) and experimental data: effect of temperature and ion valence. Symbols: experimental data for KF, CaCl$_2$ and LaCl$_3$ at 25$\,\si{\degreeCelsius}$ (\textbf{A--C}) and 55$\,\si{\degreeCelsius}$ (\textbf{D}), reproduced from Refs. \cite{goldsack1978viscosity,isono1984density}. Solid line: this work, equation \eqref{eqn:Main}. Dashed line: Falkenhagen limiting law, equation \eqref{eqn:Falkenhagen}.}
\end{figure*}

\subsection{Comparison with experimental data}

In order to assert the validity of our model, we compare our main result \eqref{eqn:Main} (and its equivalent for multivalent salts, see equation \eqref{eqn:Multivalent}) to experimental data accessible in the literature. We mainly used the data collected by Isono \cite{isono1984density}, who systematically reports the viscosity of a wide variety of electrolytes at temperatures ranging from $15 \, \si{\degreeCelsius}$ to $55 \,\si{\degreeCelsius}$ and for concentrations between $0.05 \, \si{M}$ and the saturation limit. His data unfortunately do not contain viscosity values at very high dilution, so that the Falkenhagen regime $\Delta \eta \propto c^{1/2}$ is often difficult to observe (see Fig. 2 for example). The validity of the Falkenhagen limiting law at low concentration, however, has been discussed elsewhere \cite{cox1934viscosity,anderson1994debye}. In addition, we compared Isono's data to experimental results by other experimentalists\cite{abdulagatov2006experimental,campbell1953conductances,goldsack1978viscosity,out1980viscosity}. No difference was found between the different tested datasets, and nearly all compared favorably to our model. 

Overall, we tested equation \eqref{eqn:Main} against data for the viscosity of the following salts: NaF, NaCl, NaBr, NaNO$_3$, KF, KCl, KBr, KNO$_3$, AgNO$_3$, LiCl, CaCl$_2$, MgCl$_2$, BaCl$_2$, SrCl$_2$, LaCl$_3$, Na$_2$SO$_4$ and Cd(NO$_3$)$_2$.

Equation \eqref{eqn:Main} contains three parameters that need to be specified: the Jones-Dole coefficient $B(T)$, the ion diffusion coefficient $D$, and the short-distance cut-off $a$. 

{For the diffusion coefficient, we used tabulated values at infinite dilution (see Table 1). Since its values are in general different for cations and anions, we used equation \eqref{eqn:DiffCoef} to determine the value of the effective diffusion coefficient of the electrolyte. It should be noted that the diffusion coefficients of electrolytes are found to also depend on salt concentration \cite{vitagliano1956diffusion}; however, these observations are obtained for a coarse-grained definition of the diffusion coefficient. Since we are interested here in the microscopic dynamics of ions, we assume that at the single-ion level one may use the limit of infinite dilution.

Since $a$ can be thought off as a minimum approach distance between two ions (see Fig. 4A), Avni and coworkers suggested to set $a$ to the sum of the two ionic radii (which can be determined from crystallographic data, for example). This choice, however, compares poorly to experimental data in our case. We found better agreement for higher values of $a$, which are more in line with hydrated diameter. As there is in addition an uncertainty on the exact value of $a$ (see Section \ref{SecIIF}), we used $a$ as a fitting parameter independent of temperature (see Table 2).

The Jones-Dole coefficient B(T) was determined for each temperature by examining experimental data for low concentrations.

\begin{table}
	{
	\centering
	\begin{tabular}{lc}
		Ion  & $D$ ($10^{-9}\, \si{m^2/s}$) \\
		\hline
		\hline 
		Na$^+$ & 1.33  \\
		K$^+$ & 1.96  \\
		Li$^+$ &1.03  \\
		Ag$^+$$_3$ & 1.65 \\
		Ca$^{2+}$ & 0.79 \\
		Mg$^{2+}$ & 0.705 \\
		Ba$^{2+}$ & 0.848  \\
		\hline
	\end{tabular}
	\begin{tabular}{lc}
	Ion  & $D$ ($10^{-9}\, \si{m^2/s}$) \\
		\hline
		\hline 
		Sr$^{2+}$ & 0.794  \\
Cd$^{2+}$ & 0.717  \\
La$^{3+}$ & 0.629  \\
Cl${^-}$ & 2.03 \\
NO$_3{^-}$ & 1.9 \\
SO${_4^{2-}}$ & 1.07 \\
Br${^-}$ & 2.02 \\
		\hline
	\end{tabular}
	\caption{Diffusion coefficients of ions at $25\, \si{\degreeCelsius}$.}
}
\end{table}

Lastly, note that $\lB$ itself depends on $T$, both directly through equation \eqref{eqn:BjerrumLength} and indirectly through the dielectric constant of water $\epsilon(T)$, which we determined from tabulated data \cite{malmberg1956dielectric}. Like the diffusion coefficient, $\epsilon(T)$ is known to depend on salt concentration when measured over macroscopic samples; but since again we use it here to describe the properties of water at the microscopic level around individual ions, we used the value in absence of salt. Taking this effect into account would amount to describing ion-water interactions and the hydration shell around ions; these are already encapsulated into $B(T)$ and $a$, respectively.
}

\begin{table}
	{
	\centering
	\begin{tabular}{lcc}
		Salt  & $a$ ($\si{\angstrom}$) & Ref.\\
		\hline
		\hline 
		NaCl & 5.5 & \cite{out1980viscosity} \\
		NaBr &8.5 & \cite{goldsack1978viscosity,isono1984density} \\
		NaNO$_3$ & 9.0& \cite{isono1984density}\\
		KNO$_3$ & 9.4& \cite{isono1984density}\\
		KF & 6.5 & \cite{out1980viscosity}\\
		KBr & 5.0& \cite{out1980viscosity} \\
		KCl & 6.5 & \cite{out1980viscosity} \\
		LiCl & 6.5 & \cite{out1980viscosity,abdulagatov2006experimental}\\
		\hline
	\end{tabular}
	\begin{tabular}{lcc}
	Salt  & $a$ ($\si{\angstrom}$) & Ref.\\
	\hline
	\hline 
	AgNO$_3$ & 6.5 & \cite{campbell1953conductances}\\
	CaCl$_2$ & 7.0 & \cite{isono1984density}\\
	MgCl$_2$ & 8.0 & \cite{isono1984density}\\
	BaCl$_2$ & 5.0 & \cite{isono1984density,jones1929viscosity}\\
	SrCl$_2$ & 8.5 & \cite{isono1984density}\\
	LaCl$_3$  & 6.0 & \cite{isono1984density}\\
	Na$_2$SO$_4$ & 8.5 & \cite{isono1984density}\\
	Cd(NO$_3$)$_2$ & 7.5 & \cite{isono1984density}\\
	\hline
\end{tabular}
	\caption{Fitted values of $a$ and original papers of experimental datasets for the studied salts.}
}
\end{table}

\subsection{Results and discussion}

{
Overall, we observed a very good agreement between experiments and the model, see Fig. 5. Equation \eqref{eqn:Main} quantitatively matched with literature data nearly up to the saturation limit, at all tested temperarures (see Fig. 5A-C), even for multivalent salts like BaCl$_2$ or LaCl$_3$ (see Fig. 5C and D).

In most cases, the liquid's viscosity increases with salt concentration. This is not the case for certains salts, like KCl, KBr or KNO$_3$ -- the viscosity decreases in certain concentration and temperature ranges. This effect, which is more pronounced for salts with large, weakly charged anions, is thought to be caused by \rev{interactions between ions and water molecules}. It corresponds to negative values of the Jones-Dole coefficient $B(T)$, which \rev{are linked to changes in the immediate environment of ions}. In particular, salts that decrease the liquid's viscosity tend to be those with low hydration enthalpies (for example, KCl, KBr and KNO$_3$ all have hydration enthalpies below $700 \, \si{kJ/mol}$ \cite{smith1977ionic}).

Our model is not able to predict which salt should result in a negative viscosity increment, since it does not provide a prediction for the $B$ coefficient. The model, however, correctly predicts that the viscosity should increase sharply at high concentration, see Fig. 5F in the case of KNO$_3$. In particular, the model provides a theoretical justification for the inclusion of an additional term in the Jones-Dole equation, with a $c^{3/2}$ scaling -- other usual fitting ansatz often lack theoretical ground.

The only deviation between the model and experimental data was observed when the Jones-Dole coefficient $B(T)$ changed sign. While good agreement was obtained for KCl, KBr and KNO$_3$ at low temperature (where $B(T) <0$ for these salts), the model compared poorly to experimental data for higher temperatures, where, $B(T) >0$. As this was only observed for salts for which $B(T)$ changes sign, we can suggest a modification in the hydration shell of the ions to be at the source of this effect, for example. In all other cases, agreement was good in the entire temperature range.

Lastly, we observe that the fitted values of $a$ are for the most part well above ionic radii of the corresponding salts. It should be noted that the ``correct'' way of defining the ionic size depends on context; it seems that here one should consider hydrated ions (with typical hydrated radii around $3-4 \, \si{\angstrom}$, corresponding to $a \sim 6-8 \, \si{\angstrom}$).
}

\section{Conclusion}

In this work, we derived a theoretical model for the viscosity of concentrated electrolytes. Through the use of a field-theoretical framework based on the Dean-Kawasaki equation, we recovered and considerably extended the long-standing Falkenhagen limiting law, providing a first theoretical insight on the matter beyond the limit of infinite dilution. We showed that fluctuations of charge result in an increase in viscous dissipation, scaling as the salt concentration to the power 1.5 at high concentrations, in contrast with the traditional Jones-Dole equation and similar empirical laws, but in excellent agreement with experimental data.

More importantly, we derived a general relation linking the liquid's microscopic structure factor to a macroscopic parameter like viscosity:
\begin{widetext}
{
	\begin{equation}
		\Delta \eta_\ii  =  \frac{c }{15(2 \pi)^2}\frac{\kB T}{D} \int \di  q  V( q) \deriv{}{q} \left[ q^2 C_\rho^0(q) \dpart{C_\rho^0}{q} \right].
	\end{equation}
}
\end{widetext}
This result holds in principle regardless of the precise shapes of the charge structure factor $C^0_\rho$ or the interaction potential $V$, and would be relevant in other contexts.

The conclusions of our work are two-fold. First, it shows the usefulness of the Dean-Kawasaki framework in establishing fluctuation-dissipation relationships in complex contexts, such as concentrated electrolytes. Indeed, direct computations of viscous forces due to electrostatic interactions, in line with Falkenhagen's historical derivation, are particularly arduous and therefore only tractable in simple cases, like that of infinite dilution. On the contrary, our approach allows to link quantities like the liquid's viscosity to the charge structure factor, a more easily-accessible quantity in the theory, but also in simulations or experiments \cite{kalcher2009structure,kim2023electrical,kunz1992charge}. This observation suggests manyfold potential extensions, e.g. by considering the effect of charge or density fluctuations in the solvent as well, or ion transport in more complex environments. {In particular, accounting for charge fluctuations in the solvent could allow to shed light on the effect of ion-water interactions. Another important extension would be to study the effect of solid surfaces.} In particular, the presence of surface charges typically results in a local increase in the ion concentration near walls, which could affect the viscosity of electrolytes e.g. in nanometric confinement found in nanofluidic apparatus or in biological membranes\cite{bocquet2010nanofluidics}.

Secondly, our somewhat formal and general framework does allow to catch a glimpse of the complexity and the non-universality of ion transport, by allowing to differentiate the behavior of salts with various chemical composition. While we are at this stage unable to fully rationalize specific deviations that certain salts display, we expect that this work will help to further the understanding of ion transport at the nanoscale.

\begin{acknowledgments}
	The author thanks Lydéric Bocquet, Baptiste Coquinot and Mathieu Lizée for fruitful discussions. This project has received funding from the European Union’s Horizon 2020 research and innovation programme under the Marie Skłodowska-Curie grant agreement No 101034413.
\end{acknowledgments}

\section*{Data Availability Statement}

Data sharing is not applicable to this article as no new data were created. Data supporting the findings of this study are accessible in the literature, as indicated throughout the article.

\bibliography{draft_visco.bib}

\end{document}